# Stereodirectional Photodynamics: Experimental and Theoretical Perspectives

Federico Palazzetti[a,*], Andrea Lombardi[a], Shiun-Jr. Yang[b], Masaaki Nakamura[b], Toshio Kasai[b], King-Chuen Lin[b], Dock-Chil Che[c], Po-Yu Tsai[d]

[a]*Università degli Studi di Perugia, Dipartimento di Chimica, Biologia e Biotecnologie, via Elce di Sotto 8, 06123 Perugia, Italy*
[b]*National Taiwan University, Department of Chemistry, Roosevelt Road, 10617, Taipei, Taiwan.*
[c]*Osaka University, Department of Chemistry, Toyonaka, Osaka, Japan.*
[d]*National Chung Hsing University, Taichung, Taiwan.*
*Corresponding author: federico.palazzetti@unipg.it*

**Abstract.** Hexapole oriented 2-bromobutane is photodissociated and detected by a slice-ion-imaging technique at 234 nm. The laser wavelength corresponds to the C – Br bond breaking with emission of a Br atom fragment in two accessible fine-structure states: the ground state Br ($^2P_{3/2}$) and the excited state Br ($^2P_{1/2}$), both observable separately by resonance-enhanced multiphoton ionization (REMPI). Orientation is evaluated by time-of-flight measurements combined with slice-ion-imaging.

## 1. INTRODUCTION

Advanced imaging techniques open the way to the comprehension of the dynamics of photoinitiated processes. The available energy makes possible dissociative channels, alternative to the minimum energy path, involving excited electronic states, yielding products with a large variety of energy distribution among translational and internal degrees of freedom. Our recent studies were devoted to the photodissociation dynamics of methyl formate[2-5], the simplest of esters and considered archetypal of organic molecules, with interest in astrochemistry[6] and prebiotic chemistry.[7] (For characterization of potential surface of chiral molecules see references [8-10] and for molecular dynamics simulation of enantiomeric discrimination see reference [11]). In this paper, we report the study on the photodissociation of the oriented chiral molecule 2-bromobutane at 234 nm laser wavelength. The orientational control of molecules permits to obtain information on the geometrical features and spatial aspects of dynamics otherwise concealed in free rotating ones; separation of enantiomeric effects in photodissociation and in collisional processes are among the most intriguing features expected to be associated to the molecular orientation. Here, the orientation of this asymmetric-top molecule is achieved by a 70 cm-length hexapole combined with homogeneous electric fields. An accompanying paper reports the orientation of 2-bromobutane through hexapolar technique[12]. (For details on the hexapolar technique see also references [13-20]).

The slice-ion-imaging technique is employed to study the photodissociation of the oriented 2-bromobutane with breaking of the C – Br bond at 234 nm laser wavelength, as mentioned above. The dissociated Br atom presents two accessible fine-structure states: the ground state Br ($^2P_{3/2}$) indicated by Br and the excited state Br ($^2P_{1/2}$) indicated by Br*. Both photofragments can be observed separately by resonance-enhanced multiphoton ionization (REMPI). Velocity and angular distributions have been characterized, revealing contributions from the two isotopes of bromine atom, $^{79}$Br and $^{81}$Br, which have similar abundance in nature; the use of weak ion extraction field allows one to separate the two contribution to the signal. Angular distribution of photofragments can be predicted by vector correlation of intrinsic molecular properties (*e. g.* permanent dipole moment and transition dipole moment) and external vectors defined in the laboratory-frame (*e. g.* laser polarizations and orientation vector). Evaluation of the molecular orientation has been obtained through (i) time-of-flight measurements: the arrival time depends on the forward or backward displacement of the bromine atom; (ii) through the top-down asymmetry detected in slice ion imaging.

## 2. HIGHLIGHTS ON THE EXPERIMENTAL APPARATUS

The experimental apparatus has been described in detail previously[21]. The vacuum system consists of three chambers and contains the molecular-beam source, the electrostatic hexapole and the imaging detector system, separated by differential pumping. After passing the hexapole chamber, the molecular beam reaches the ion lens, where crosses perpendicularly a linear polarized laser. The ion lens is composed of four hollow metal disks, according to the "three-electrode" design (see reference [21] and references therein). The sliced images are centered on $^{79}$Br. The ion lens generates a static electric field of 200 V/cm that works as orienting field and as ion extractor, accelerating the photofragments to the detector. A pulsed voltage added in the extraction stage permits to change intensity and sense of the orienting field. The pulsed voltage is applied when the molecular beam reaches the repeller plate and lasts 28 μs, stopping before the intersection with the linearly polarized laser. Being the response time of the pulsed voltage very short, it takes only 30 ns and it does not interfere with the ion extraction.

## 3. PHOTODISSOCIATION

Photodissociation of the hexapole oriented 2-bromobutane has been carried out at a laser wavelength of *ca.* 234 nm, the Br and Br* fragments have been ionized at 233.7 and 234.0 nm, respectively, by (2+1) REMPI. Velocities and angular distributions of the dissociated bromine atoms have been obtained by sliced ion images measurements (Figure 1). The maximum velocity is 965 m/s for Br and 920 m/s for Br*, which corresponds to a center-of-mass total translational energy disposal of 87.8 kJ/mol for Br and 79.8 kJ/mol; the anisotropy parameter $\beta$ is 1.49 for Br and 1.85 for Br*. This study permitted to amend a previous investigation on the photodissociation of 2-bromobutane,[22] establishing the contribution of the natural isotopes of bromine, $^{79}$Br and $^{81}$Br, to the velocity distribution. An in-depth measurement, performed by a series of time slices of both isotopes by shifting the pulsed gate timing, permitted to attribute an inner region of the image, not observed in our experiment, to $^{81}$Br.

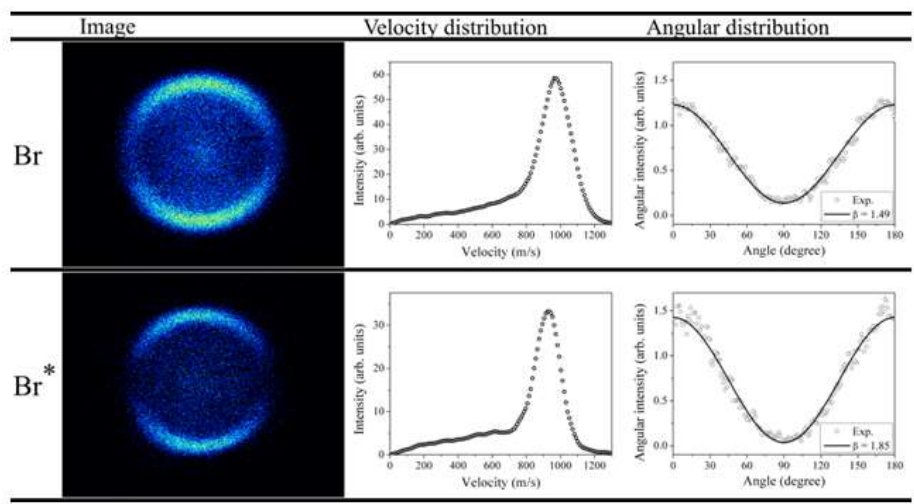

**FIGURE 1.** The sliced images (the gate pulse width is 25 ns) of Br and Br* fragments obtained by photodissociation of 2-bromobutane at 233.7 nm and 234.0 nm, respectively. Related velocity and angular distributions are in the central panel.

Velocity map imaging of the with a gate width of 400 ns is reported in Figure 2. With respect to the results shown in Figure 1, where the gate width is 30 ns, it is possible to cover the whole ion cloud spheres, both for $^{79}$Br and $^{81}$Br. In Figure 2, we report the experimental (2a) and simulated (2b) imaging and the angular distribution of the photofragment (2c). This latter has been fit by the equation

$$I(\theta) = 1 + \beta_1 P_1(\cos\theta) + \beta_2 P_2(\cos\theta),$$

where $P_1$ and $P_2$ are terms of the Legendre polynomials, and $\beta_1$ and $\beta_2$ are the expansion coefficients. More precisely, $\beta_1$ is related to the top-down asymmetry of the angular distribution (the ring in the image) and is non zero only for oriented molecules[23]; for Br*, $β_1$ is 0.53, while $β_2$ is 1.56. Correlation of the recoil velocity vector of the photofragment and the permanent and transition dipole moment vectors of the photodissociated molecule permit to characterize the spatial fragment distribution. These vectors have been inferred from the simulated 2D projection of the three-dimensional photofragment angular distribution (Fig. 2b) (details on the vector correlation of the photodissociation of the oriented 2-bromobutane are given in Nakamura *et al.* [21]).

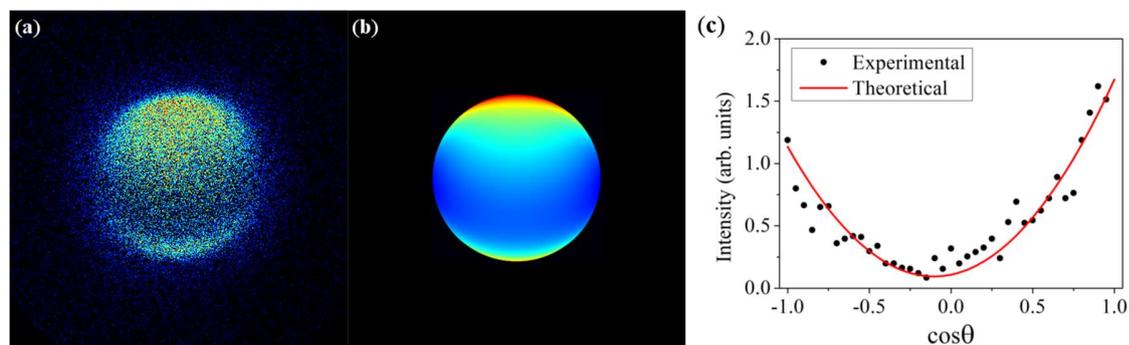

**FIGURE 2.** The experimental (a) and simulated (b) ion image (with a 400 ns long gate width) of Br* photofragment. (c) Comparison between experimental and simulated angular distribution of Br* photofragment.

## 4. CONCLUSIONS

Control of the molecular orientation is a crucial point in the study of stereodynamics of elementary processes. For the first time, a complex molecule such as 2-bromobutane has been oriented by the electrostatic hexapolar technique and the photodissociation dynamics has been investigated by slice ion imaging. A pulsed voltage applied at the ion extraction stage permitted to control the intensity and sense of the orienting field. Angular and velocity distributions of Br ($^2P_{3/2}$), the ground state, and Br ($^2P_{1/2}$), the first excited state, has been measured and the spatial distribution of Br ($^2P_{1/2}$) have been calculated through the vector correlation model.

Future investigations will involve photodissociations among other excited states. The use of the pulsed voltage, which permits to separate the ion extraction and orienting stages, avoids interferences between these two stages allowing to reach higher orienting field intensities than those commonly used, in this case 200 V/cm, making the technique suitable even for more complex molecules. Investigations of the separated enantiomers are expected to contribute to the important issue of the photochemical control of chirality.

## ACKNOWLEDGMENTS

FP, AL and VA acknowledge the Italian Ministry for Education, University and Research (MIUR) for financial support ("Scientific Independence for young Researchers", SIR 2014, RBSI3U3VF).